\documentclass[draft]{agujournal2019}
\usepackage{url} %this package should fix any errors with URLs in refs.
\usepackage[inline]{trackchanges} %for better track changes. finalnew option will compile document with changes incorporated.
\draftfalse

\journalname{Space Weather}

\begin{document}

\title{Forecasting $>$300\,MeV SEP events: Extending SPARX to high energies}

\authors{C.O.G. Waterfall\affil{1}, S. Dalla\affil{1}, M.S. Marsh\affil{2}, T. Laitinen\affil{1}, A. Hutchinson\affil{1}}

\affiliation{1}{University of Central Lancashire}
\affiliation{2}{Met Office}

%1) Aboa Space Research Oy, Tierankatu 4B, 20520 Turku, Finland
%2) Department of Physics and Astronomy, University of Turku, 20014 Turku, Finland

\affiliation{1}{Jeremiah Horrocks Institute, University of Central Lancashire, Preston, UK, PR1 2HE}
\affiliation{2}{Met Office, Exeter, UK}

\correspondingauthor{Charlotte Waterfall}{cwaterfall@uclan.ac.uk}

\begin{keypoints}
\item A new version of the SPARX forecasting model is presented, extended to high energies above 300\,MeV
\item Correlations between peak X-ray flare magnitudes and peak proton fluxes are improved when events are grouped by their IMF polarity
\item For a given flare magnitude, events that occur during an A$-$ polarity have larger peak fluxes than for A+ events
%correlations improve when events separated according to polarity
%new forecasting model for high energies
\end{keypoints}

\begin{abstract}
The forecasting of solar energetic particles (SEPs) is a prominent area of space weather research. Numerous forecasting models exist that predict SEP event properties at proton energies $<$100\,MeV. One of these models is the SPARX system, a physics-based forecasting tool that calculates $>$10\,MeV and $>$60\,MeV flux profiles within minutes of a flare being detected. This work describes SPARX$-$H, the extension of SPARX to forecast SEP events above 300\,MeV . SPARX$-$H predicts fluxes in three high energy channels up to several hundred MeV. Correlations between SEP peak flux and peak intensity of the associated solar flare are seen to be weak at high energies, but improved when events are grouped based on the field polarity during the event. Initial results from this new high energy forecasting tool are presented here and the applications of high energy forecasts are discussed. Additionally, the new high energy version of SPARX is tested on a set of historic SEP events. We see that SPARX$-$H performs best when predicting peak fluxes from events with source locations in well-connected regions, where many large SEP events tend to originate.
\end{abstract}

\section*{Plain Language Summary}
High energy particles, accelerated in eruptions from the Sun, can quickly travel throughout space. If they reach Earth, these particles pose a hazard to technology and life, disrupting environments from satellites in orbit to aircrew and passengers at altitude.  The highest energy solar particles are rarer, but they can reach Earth in timescales as short as minutes. Therefore, forecasting models are needed to evaluate their strength, duration, and other features. This work describes a new version of a solar energetic particle forecasting model. An existing model is extended to operate at high energies, predicting properties of the particle event that can be used to mitigate associated hazards. The new model is tested against previous particle events and the outputs are compared to observed features. The model produces a forecast for 72$\%$ of previously observed particle events.

%\keywords{Solar Energetic Particles (SEP) -- Heliosphere -- Sun -- Forecasting -- Space weather  }

%   \titlerunning{Extending SPARX to $>$350\,MeV}

 %% context heading (optional). leave {} empty if necessary

%\maketitle

\section{Introduction}
The acceleration and propagation of solar energetic particles (SEPs) throughout interplanetary space are a significant hazard to technology and manned spaceflight. The consequences of higher energy SEP events (up to several GeV) can have far reaching consequences including significant risks to technology (e.g. satellites and avionics), as well as biological risks to those at high altitude (aviation crew and passengers) or in space and are therefore crucial to forecasting models. Forecasts at relativistic energies are needed to help inform us of what steps need to be taken, e.g. when or how long to put electronics into safe mode during a storm, when to bring astronauts inside during an extravehicular activity (EVA), whether to divert an aeroplane, etc. With the ongoing rise in space exploration there comes the need for more reliable forecasting models to mitigate the radiation risks following a solar eruptive event. Fortunately, with this growing need there has also been an increase in the number of SEP forecasting models developed. There are a wide variety of existing forecasting models; some employing physics-based calculations (e.g. \citeA{Luhmann2007,marsh2015sparx,Wijsen2020PhD}), others taking an empirical approach (e.g. \citeA{Anastasiadis2017,Richardson2018,bruno2021empirical}). Additionally, some models provide forecasts of the entire flux profile of the event, while others predict singular event properties, e.g. peak flux. An in-depth review of 35 existing SEP forecasting models is given in \citeA{whitman2022review}.

Many of the forecasting models outlined in \citeA{whitman2022review} predict flux properties for proton energies around 10\,MeV proton events and generally it is rare for forecasting models to consider protons at energies $>$100\,MeV. Exceptions include: Solar Energetic Particle MODel (SEPMOD) \cite{luhmann2007heliospheric}, FOrecasting Solar Particle Events and Flares (FORSPEF) \cite{anastasiadis2017predicting} etc. This paper details progress in the forecasting of relativistic protons by extending an existing SEP forecasting model, SPARX (Solar Particle Radiation SWx, \citeA{marsh2015sparx}), to energies above 300\,MeV as part of the UK's Space Weather Instrumentation, Measurement, Modelling and Risk (SWIMMR) programme.

SPARX is a physics-based solar energetic particle (SEP) forecasting model that produces SEP time-flux profiles following a solar eruptive event. It is capable of producing a forecast SEP profile within minutes of being triggered by the observation of a solar flare. The model is based upon results of a 3D test particle code and includes drift effects in a unipolar magnetic field due to the curvature and gradient of the Parker spiral. These drift effects are especially important for high energy particles, as their propagation can be strongly affected by the interplanetary magnetic field polarity \cite{dalla2013solar,dalla20203d}. The particles are injected instantaneously at 2R$_\odot$ (delta function in time) at a broad source region centered on the flare location, representing a CME-driven coronal shock, and allowed to propagate through the heliosphere in a Parker spiral magnetic field. No secondary injection, i.e. from a shock at interplanetary locations, is modelled. The acceleration mechanism, e.g. magnetic reconnection in solar flares or shocks driven by coronal mass ejections (CMEs) \cite{klecker2006energetic}, is also not modelled.

Originally, SPARX was designed to forecast protons with energies of $>$10\,MeV and $>$60\,MeV. This paper presents an extension of SPARX to forecast higher energy (above 300\,MeV) SEP profiles in what will be called SPARX$-$H here. The methodology used and the first results from the new high energy version are discussed. There are several similarities and differences between the two versions of SPARX which will be highlighted in this paper. A full description of the original version of SPARX and of the forecasting approach can be found in the previous work by \citeA{marsh2015sparx}. SPARX$-$H is being developed as part of the SWIMMR Aviation Risk Modelling (SWARM) project and is motivated by the need to forecast radiation during ground level enhancements (GLEs), the largest SEP events.

%sentence could be moved earlier in generic statement about space weather. radiation risk for astronauts during EVA could be added

Section~\ref{sec:extending} outlines how SPARX has been modified to forecast $>$300\,MeV protons whilst still maintaining the original functionality. This includes a new analysis of high energy SEP events as observed by High Energy Proton and Alpha Detector (HEPAD) onboard the Geostationary Operational Environmental Satellites (GOES) between 1984$-$2017. Section~\ref{sec:results} provides an example output from SPARX$-$H, as well as an analysis of the predicted properties of a selection of historic test events. Comparisons with SPARX are also made and SPARX$-$H's relative strengths and weaknesses are outlined.

%\begin{itemize}
%    \item introduction
%    \item sep forecasting models
%    \item why high energy ones are necessary
%    \item low energy SPARX
%    \item layout of paper
%\end{itemize}

\section{Extending SPARX to $>$300\,MeV}
\label{sec:extending}
The SPARX system is comprised of two primary elements; a pre-generated database of test particle model runs and the real-time SPARX tool that forecasts flux profiles following the detection of an $>$M1.0 flare. A detailed description of the 3D test particle code that is used to generate the database is given in \citeA{marsh2013drift}. Each run in the database describes an injection of protons at a different location near the Sun. When a flare is detected in real-time, its location and magnitude are input into SPARX which pulls the relevant test particle runs from the database to build a flux profile. CME parameters are not used since they are not currently available in real-time. Using a pre-generated database allows for the computation time from triggering to output to be only a few minutes, compared to typical run times of the 3D test particle code of several hours. However, in order for this rapid computation time the database is generated for a fixed set of parameters. For example, a pitch angle scattering mean free path of $\lambda$=0.3\,AU is assumed, representative of values derived from relativistic proton event data \cite{bieber2002energetic,bieber2004spaceship} (and also used in the original SPARX).

One of the parameters kept constant in the database is the energy range of the injected protons. In SPARX, protons are injected according to a power law in energy with a spectral index of $\gamma$=1.1. For the original SPARX, the proton energy range covered is 10$-$400\,MeV. As this maximum energy is too low for our intended high energy version, the entire database was regenerated for particle energies up to 1200\,MeV. As SPARX$-$H is designed as an extension to SPARX rather than an entirely new model, every other parameter was kept the same. Further details on the specific parameters can be found in the original paper \cite{marsh2015sparx}.

The other key element that required modification was the scaling of the model count rates into physical flux units. In SPARX this was done by utilising relationships between peak GOES soft X-ray fluxes (i.e. flare magnitudes) and peak proton fluxes observed during SEP events. These relationships were obtained from a prior study of SEP events and their associated flare and CME parameters by \citeA{dierckxsens2015relationship}. In SPARX, the average proton peak flux in five different flare intensity bins from M1.0 to $>$X5.0 is taken from \citeA{dierckxsens2015relationship} and used to scale the simulated fluxes. This was done for the integral energy channels $>$10 and $>$60\,MeV in order to compare with standard GOES SEP data. As we are considering energies corresponding to the largest SEP events (i.e. over several hundred MeV), a new analysis of high energy SEP events was required to acquire the relevant scaling.

In order to extend SPARX we had to first perform an analysis of $>$300\,MeV proton events. The data used was obtained from observations by the GOES series of spacecrafts between 1984$-$2017, which include high energy particle events detected by HEPAD. The GOES$-$HEPAD dataset remains the most comprehensive high energy dataset that has the least data gaps and covers the largest time period. We used a newly calibrated, background subtracted and cleaned version of this dataset, developed as part of the ESA HIERRAS project. The dataset is based on the SEPEM reference data set (version 2; RDS v2) \cite{jiggens2012,jiggens2018}, extended in energy with re-calibrated HEPAD data \cite{raukunen2020very}. The new nominal high energy channels are listed in Table~\ref{table.channels}. 42 SEP events from this dataset were identified that were associated with a solar eruption between longitude E90 and W90. The number of SEP events are far fewer in number than those studied by \citeA{dierckxsens2015relationship}, a consequence of the relative rarity of $>$300\,MeV events compared to their lower energy counterparts. A full table of events and an in-depth analysis of these high energy events can be found in \citeA{waterfall2022hepad}.

 \begin{table}
 \caption{Proton energy channels of the re-calibrated GOES dataset \cite{raukunen2020very}[cm$^{-2}$s$^{-1}$sr$^{-1}$MeV$^{-1}$].}
 \centering
 \begin{tabular}{c c c}
 \hline
 \textbf{Channel} & \textbf{Central energy [MeV]} & \textbf{Energy range [MeV]} \\
 \hline
Ch12 & 347.8  & 289.2--418.3 \\
Ch13 & 503.0   & 418.3--604.9 \\
Ch14 & 727.4 & 604.9--874.7 \\
 \hline
 %\multicolumn{2}{l}{$^{a}$Footnote text here.}
 \end{tabular}
 \label{table.channels}
 \end{table}

 %The dataset used in this study was constructed within the ESA HIERRAS project. It is based on the SEPEM reference data set (version 2; RDS v2) \cite{jiggens2012,jiggens2018}, extended in energy with re-calibrated HEPAD data \cite{raukunen2020very}. The RDS v2 is based on proton data observed by the SMS and GOES satellites which has been cleaned and cross-calibrated using IMP-8/GME data \cite{sandberg2014,rodriguez2017}. The instrument channels have been re-binned into 11 logarithmically spaced energies between 5 and 289.2 MeV.

%There have been efforts to combine all GOES$-$HEPAD observations into one calibrated and cleaned dataset \cite{raukunen2020very}. This new dataset is based on the SEPEM reference dataset (RDS version 2) and covers events between 1984$-$2017. A full analysis of that dataset is given in Waterfall et al. 2022 in prep. We chose not to use this merged dataset and carry out our scaling with the original HEPAD data here for several reasons: to keep the predicted high energy fluxes consistent with the observed energy channels (the merged data has been re-binned into different energy channels to those observed by HEPAD), to avoid the strong background subtraction on the merged dataset which has removed some SEP events (as we are already working with a reduced number due to the rarity of these events), to keep consistency between predicted flux outputs and the original GOES data over the high energy range.

\begin{figure}%[H]
\centering
\noindent\includegraphics[width=\textwidth]{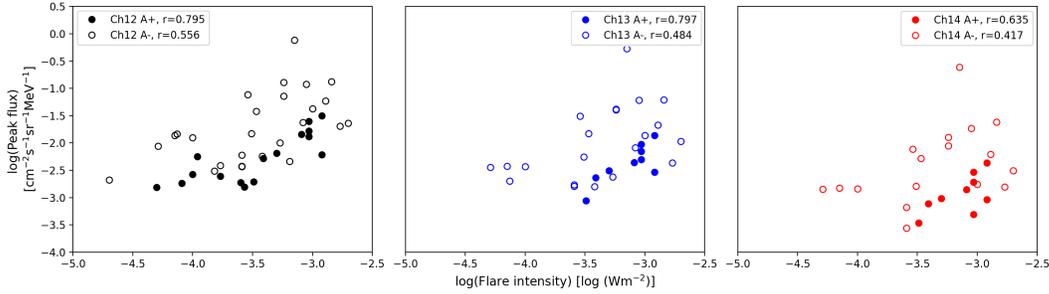}
\caption{Plot of the logarithm of the flare intensity versus logarithm of the proton peak flux for the three high energy GOES channels. The y-axes span equal orders of magnitude. A+ and A$-$ events are shown as filled and empty circles respectively. Black, blue and red circles correspond to events in Ch12, 13 and 14 respectively (left to right panels). Correlation coefficients, r, for each group are labelled.}\label{fig:peakfluxmag}
\end{figure}

As in \citeA{dierckxsens2015relationship}, \citeA{waterfall2022hepad} examined correlations of the peak flux of these events with the flare magnitude and CME speeds. \citeA{dierckxsens2015relationship} found values of the Pearson correlation coefficient of 0.55 and 0.63 between the peak flux and flare magnitude for $>$10\,MeV and $>$60\,MeV. Correlation coefficients with flare SXR peak intensity for Ch 12, 13 and 14 events are 0.55, 0.39 and 0.28 respectively. The relationships are poor, becoming weaker with increasing energy. There is no dependence on the longitude of the source region associated with these events. However, when the events are divided according to the polarity of the interplanetary magnetic field (IMF) at the time of the event these correlations improve across all energies, as shown in Figure~\ref{fig:peakfluxmag}. An A+ (A$-$) polarity describes a magnetic field configuration where open field lines mostly point outwards (inwards) in the northern hemisphere and inwards (outwards) in the southern hemisphere. The orientation is determined via lookup tables defining the A+ and A$-$ date ranges derived from manual analysis of synoptic source surface maps (SSMs) provided by the Wilcox Solar Observatory (WSO) at: \url{http://wso.stanford.edu}). These maps are created from potential field modelling using photospheric magnetogram data. For an A$-$ polarity the correlations are 0.56\,$\pm$\,0.042, 0.48\,$\pm$\,0.049 and
0.42\,$\pm$\,0.062 (Ch12, 13, 14). For an A+ polarity the corresponding coefficients are 0.80\,$\pm$\,0.020, 0.80\,$\pm$\,0.050 and 0.64\,$\pm$\,0.087. The errors are determined via bootstrapping and calculating the standard deviation from the re-sampled correlation coefficients. This grouping reveals the tendency for A+ events to have smaller peak fluxes for a given flare intensity, as well as more high energy events associated with an A$-$ polarity. The dependence of the relativistic proton transport on the polarity was discussed by \citeA{dalla20203d} and \citeA{waterfall2022modeling}.
While the relationship between peak flux and flare intensity for A+ events is generally strong, it should be noted that the number of A+ events is less A$-$ events over all energies. For Ch12, 13 and 14 respectively the number of A+ (A$-$) events is: 16 (26), 9 (21) and 9 (17).
\citeA{owens2022solar} reported that more GLE activity occurs during the early phase in even solar cycles and during the late phase in odd cycles: these two phases correspond to times of A$-$ polarity and thus their results confirm the polarity dependence of relativistic SEP events that we observed.

%The largest flare in this event set is Largest flare in our sample, smallest, median size flare, mode flare, average, largest flares ($>$X10) are all GLEs with the non-GLE SEP events peaking around X1.0.

%One difference of the higher energy analysis seen in Figure~\ref{fig:peakfluxmag}, compared to the \cite{dierckxsens2015relationship} study, is the flare magnitude range the events occupy. The events in \cite{dierckxsens2015relationship} extend down to M1.0 flares whilst the smallest flare in our sample is an M5.0. The largest is an X17 (GLE 65). The median size flare in our sample is an X3.4. The distribution of the flare sizes in our sample for SEP and GLE events is shown in Figure~\ref{fig:sparxhisto}. The largest flares ($>$X10) are all GLEs, with the non-GLE SEP events peaking around X1.0. Due to the larger flare sizes associated with high energy SEP events (and no $>$350\,MeV SEP events associated with a flare of $<$M5.0), the trigger for SPARX$-$H has changed from M1.0 to M5.0. We have not included SEP events with associated flares behind the limb ($>$W90) as the flare magnitudes are often unreliable. For example, GLE 61 (18 April 2001) was a large event (neutron monitor maximum increase $>$20$\%$) with an associated flare at W115 and small magnitude of C2.2.

%\begin{figure}%[H]
%\centering
%\noindent\includegraphics[width=0.7\textwidth]{sparxhisto.png}
%\caption{Distribution of associated flare magnitudes for our SEP events observed between 1997$-$2017. GLE events are highlighted in red, non-GLE SEP events in blue.}\label{fig:sparxhisto}
%\end{figure}

When calculating the new flare intensity bins needed for SPARX$-$H, the events are first divided based on the IMF polarity, following the stronger correlations seen between A$+$ and A$-$ evens. As there are fewer $>$300\,MeV events compared to low energies, the number of flare intensity bins that SPARX uses is reduced from 5 down to 3 for Channel 12, to ensure each bin contains a sufficient (and similar) number of events. Channels 13 and 14 have no events in the $<$X2 flare magnitude bin of the A+ group. For verification and validation purposes later (see Section~\ref{sec:results}), $\sim$30$\%$ of the events are randomly selected from each group and removed before each bin is calculated. These `test events' are later used in Section~\ref{sec:results} for testing. The correlation coefficients are not affected (within $\pm$0.03) by this. Table~\ref{table.scalings} shows the average of the logarithm of the peak fluxes in the three bins (and two polarity groups) for Ch 12, 13 and 14. Bootstrapping is again performed to find an error, randomly re-sampling and removing $\sim$30$\%$ of the events 10 times to find the standard deviation, given as the error in Table~\ref{table.scalings}. The average peak fluxes are now used in SPARX$-$H to normalise the model count rates generated by the test particle code into physical flux units.

\begin{table}
\caption{Mean of the logarithm of the proton peak flux (log [cm$^{-2}$\,s$^{-1}$\,sr$^{-1}$\,MeV$^{-1}$]) for each of the high energy channels in the three flare intensity bins used by SPARX$-$H.}
\centering
\begin{tabular}{ccccccc}
\hline
\textbf{Flare intensity} & \multicolumn{2}{c}{\textbf{Ch12}} & \multicolumn{2}{c}{\textbf{Ch13}} & \multicolumn{2}{c}{\textbf{Ch14}} \\
\textbf{} & A+ & A- & A+ & A- & A+ & A- \\ \hline
\textless{}X2 & -2.68\,$\pm$\,0.06 & -2.16\,$\pm$\,0.08 &  & -2.56\,$\pm$\,0.04 &  & -2.84$\pm$0.06 \\
X2$-$X7 & -2.43\,$\pm$\,0.05 & -1.80\,$\pm$\,0.13 & -2.57\,$\pm$\,0.11 & -2.19\,$\pm$\,0.16 & -3.06\,$\pm$\,0.09 & -2.60\,$\pm$\,0.19 \\
\textgreater{}X7 & -1.69\,$\pm$\,0.09 & -1.05\,$\pm$\,0.09 & -2.08\,$\pm$\,0.10 & -1.40\,$\pm$\,0.11 & -2.73\,$\pm$\,0.15 & -1.96\,$\pm$\,0.12 \\ \hline
\end{tabular}
\label{table.scalings}
\end{table}

The inputs of SPARX$-$H remain the same as for SPARX: flare latitude, longitude and magnitude determined from GOES soft X-ray measurements. An additional input for SPARX$-$H is added where the user specifies the date, from which the polarity of the interplanetary magnetic field is determined and used to dictate which scaling is required. The user can also manually input the polarity. The trigger for the operation of SPARX$-$H is increased from $>$M1.0 to $>$M7.0 as the majority of these higher energy events are associated with larger flares. In this event set, the median flare size is X3.6 with 86$\%$ of events associated with flares of X1.0 or higher. The outputs of SPARX$-$H are still 10 minute averaged proton flux profiles (where the averaging period can be altered within SPARX). However, the outputs are now three differential proton flux profiles corresponding to Ch 12, 13 and 14 from Table~\ref{table.channels}. SPARX also outputs information on the peak time, event duration and peak flux values. The following section shows some an example output from SPARX$-$H, as well as an analysis of how well SPARX$-$H performs. It is important to note that the SPARX$-$H flux profiles are generated in differential energy channels rather than the integral channels used in the original SPARX.

\section{SPARX$-$H outputs and testing}
\label{sec:results}

\begin{figure}%[H]
\centering
\noindent\includegraphics[width=0.8\textwidth]{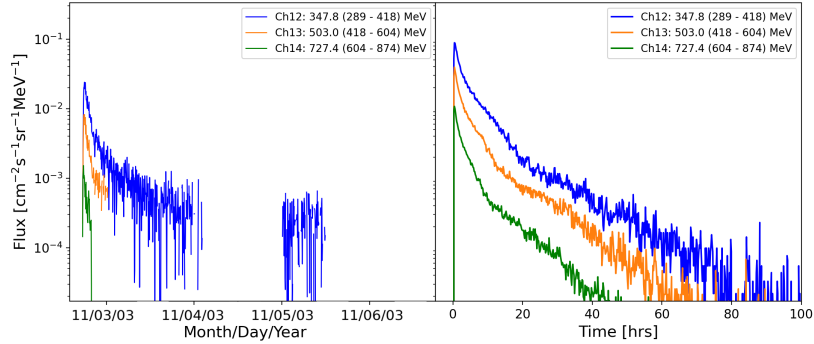}
\caption{Comparison of the forecast flux profile in Ch 12, 13 and 14 from SPARX$-$H (right) and the calibrated and background subtracted dataset from GOES observations (left) for the GLE event of 2 November 2003. Both observed and forecast profiles are shown for a period of 100 hours following the flare.}\label{fig:compare}
\end{figure}

SPARX and SPARX$-$H predict event properties including the peak flux values, time of peak flux and onset time. Another benefit of the SPARX system is its ability to provide information on the evolution of the flux and fluences in different energy channels. To test how SPARX$-$H performs, we have run SPARX$-$H for the test events randomly excluded in Section~\ref{sec:extending}. Figure~\ref{fig:compare} shows both the observations and SPARX$-$H output profiles for one of the test events: GLE 67 on 2 November 2003. This event had an associated solar flare of magnitude X8.3 located at S14W56. Both plots show three distinct impulsive profiles that decay over several hours in all three high energy channels from Table~\ref{table.channels}, where the predicted flux profiles are computed by integrating the counts at a specified 1AU observer location. As expected, the Ch12 profiles have the highest peak flux. When events with more eastern longitudes are modelled, the peak flux becomes progressively lower, with extreme eastern events producing no SEP forecast at Earth. This supports the reduced occurrence of eastern high energy SEP events observed in \citeA{waterfall2022hepad}, where only 7 of the 42 events had an eastern longitude, none greater than E64. As such, another feature of SPARX is the ability to reproduce the East-West asymmetry seen in SEP flux profiles \cite{marsh2015sparx}. Eastern events typically display a much more gradual rise phase, with the zone of good magnetic connectivity in the West producing more impulsive profiles. This effect is also seen in SPARX$-$H.

There are several quantitative features of these profiles which can be extracted for the event of Figure~\ref{fig:compare}, and the other test events, to gauge how well SPARX$-$H forecasts these historic events.

\subsection{Testing with historic SEP events: profile parameters}
We define a flux threshold for the SPARX$-$H forecasts in each channel, e.g. 1$\times$10$^{-4}$ [cm$^{-2}$s$^{-1}$ sr$^{-1}$MeV$^{-1}$] in Channel 12, based on observed event fluxes. Any forecasts which predict peak fluxes lower than this (or are poorly defined with no continuous counts lasting $>$3 hours) are not accepted as a positive SEP forecast. With this threshold set, 3 of the 13 test events do not forecast SEPs at Earth. The 3 events that fell below the threshold had longitudes $>$W75 (2 events) or had an associated A+ polarity (1 event). The remaining 10 events had clearly defined flux profiles well above the threshold. Features of the forecasts produced by SPARX$-$H for these events are now examined and compared with actual GOES measurements. A full description of how features such as event onset, duration and rise time are calculated from the calibrated dataset is given in \citeA{waterfall2022hepad}.

\subsubsection{Peak flux}
\begin{figure}
\centering
\noindent\includegraphics[width=\textwidth]{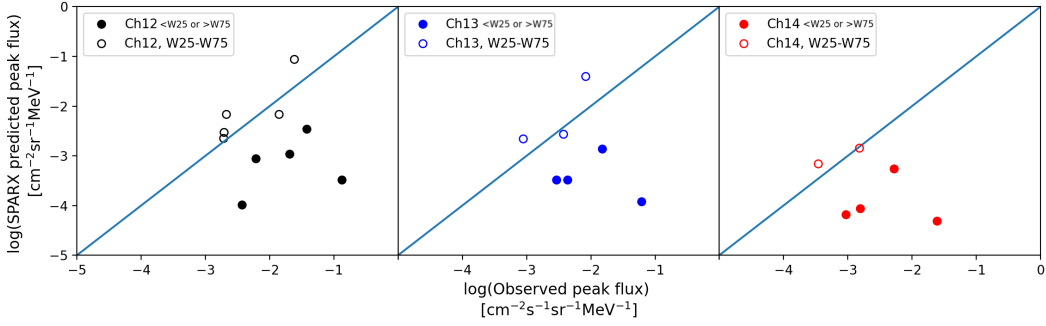}
\caption{The observed and SPARX$-$H predicted logarithm of the peak fluxes for the historic test SEP events over the three high energy channels. Unfilled circles show events with associated flare longitudes between W25 and W75. The blue line is the 1:1 line.}\label{fig:comparepeaks}
\end{figure}

Figure~\ref{fig:comparepeaks} shows the predicted peak fluxes from the historic test events across the 3 high energy channels. Generally SPARX$-$H has a tendency to under-predict the peak flux. The events which lie closest to and above the 1:1 line have well-connected source longitudes between W25$-$75 (shown as unfilled circles). The events furthest below the line have either eastern longitudes or longitudes $>$W80. The two events which fail to match observations by several orders of magnitude are the 15 April 2001 event with source longitude W85 (from \citeA{waterfall2022hepad}) and 24 September 2001 with source longitude E23. % \textbf{The Pearson correlation coefficients are given in Table~\ref{table.pearson} for the three energy channels. While the number of test events are small, there is a weak negative correlation present across the channels. This correlation improves (and becomes positive) when the poorly connected events are removed.}

 %\begin{table}
 %\caption{Pearson correlation coefficients for the three observed and predicted event parameters in the three high energy channels from Figure~\ref{fig:comparepeaks},~\ref{fig:compareduration} and~\ref{fig:comparerise}.}
 %\centering
 %\begin{tabular}{l c c c}
 %\hline
 %Event parameter &\textbf{Ch12} & \textbf{Ch13} & \textbf{Ch14} \\
 %\hline
%Peak flux, Figure~\ref{fig:comparepeaks} & -0.0199 & -0.282	& -0.408 \\
%Event duration, Figure~\ref{fig:compareduration} & 0.468 & 0.857 & -0.402 \\
%Rise time, Figure~\ref{fig:comparerise} & 0.479 & 0.369	& -0.420 \\
% \hline
 %\multicolumn{2}{l}{$^{a}$Footnote text here.}
% \end{tabular}
% \label{table.pearson}
% \end{table}

\subsubsection{Event duration}
\begin{figure}%[H]
\centering
\noindent\includegraphics[width=\textwidth]{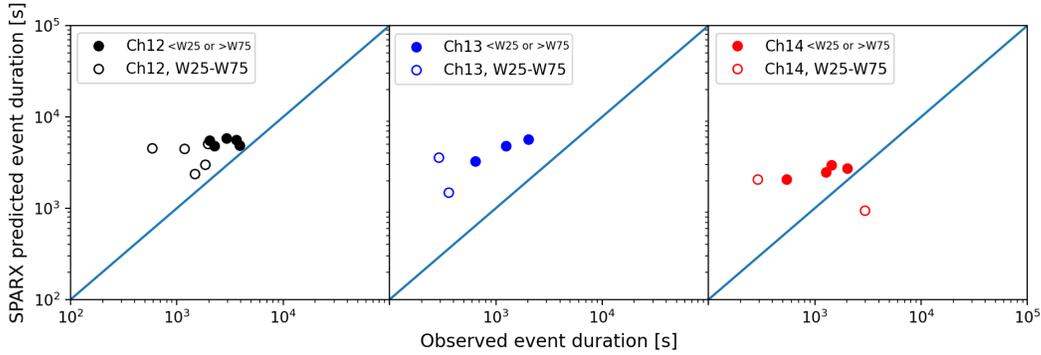}
\caption{The observed and SPARX$-$H predicted event durations, from start to end in minutes, for the historic test SEP events over the three high energy channels. The blue line is the 1:1 line. The upper limit from the predicted values originates from the simulation period of 100 hours. Unfilled circles show events with associated flare longitudes between W25 and W75.}\label{fig:compareduration}
\end{figure}

Figure~\ref{fig:compareduration} compares the observed and predicted event durations in the three high energy channels. Generally SPARX$-$H over-predicts the durations of the historic events.  This pattern remains the same over all three channels. The calibrated dataset observations are primarily obtained from HEPAD data, which is known to have a high background (which the model does not have). The strong background subtraction applied to this dataset will therefore affect the event duration calculation, removing the lower fluxes towards the end of the event. Additionally, there are two events which are part of a sequence of SEP events separated by less than 3 days. In this case, the end time of the observed event is taken to be the start of the secondary event. For example, GLE 65 (28 October 2003) is one of the test events, but it was followed by another GLE only a day later. SPARX$-$H currently only forecasts single events so there is no limit to the event duration from this.  The time when the flux passes the event threshold in each channel is taken as the start of the event. There is an upper limit to the event durations predicted by SPARX$-$H due to the test particle runs having a simulation period of 100 hours, however, none of the test events considered here have an observed event duration greater than 100 hours.

The unfilled circles denote those events with associated flare longitudes between W25$-$W75, as in Figure~\ref{fig:comparepeaks}. There is less dependency on the source location here, with the events that are the least well predicted having the smallest associated flare magnitudes, $<$M7.  %The Pearson correlation coefficients for the predicted and observed event durations are given in Table~\ref{table.pearson}. There are reasonable positive correlations for Ch12 and 13. For Ch 14 however, this correlation becomes negative due to the two well connected events with the smallest (M class) associated flare magnitudes.}

%\begin{figure}%[H]
%\centering
%\noindent\includegraphics[width=\textwidth]{decaytimes.png}
%\caption{decay}\label{fig:comparedecay}
%\end{figure}

\subsubsection{Rise time}
\begin{figure}%[H]
\centering
\noindent\includegraphics[width=\textwidth]{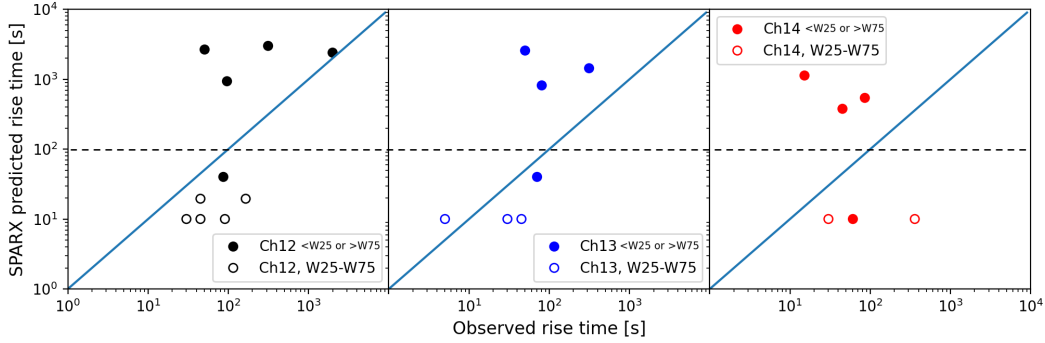}
\caption{The observed and SPARX$-$H predicted rise times in minutes for the historic test SEP events over the three high energy channels. The blue line is the 1:1 line, the dashed line indicates a predicted rise time of 100 minutes, below which are all well-connected Western events. Unfilled circles show events with associated flare longitudes between W25 and W75.}\label{fig:comparerise}
\end{figure}

Finally, the observed and predicted rise times are shown in Figure~\ref{fig:comparerise}. For all three energy channels there are two distinct groups. Those with short predicted rise times (e.g. less than 100 minutes, as marked by the dashed line) are all well-connected events with longitudes between W25 and W75 (denoted by the unfilled circles).  The filled circle with a rise time less than 100 minutes is the 13 December 2006 SEP event which had a source longitude of W23 (from \citeA{waterfall2022hepad}). Above the dashed line, with predicted rise times much greater than 100 minutes are events with source longitudes $<$W23 or $>$W75. Of these events, the worst SPARX$-$H predictions are for those with the least well-connected longitudes, i.e. Eastern or $>$W80. This feature of long rise times for forecast eastern events reflects the gradual nature of observed eastern flux profiles. %\textbf{The correlation coefficients for the three high energy channels again show weakly positive correlations for Ch12 and 13, becoming negative for Ch14. The events which are poorly predicted by SPARX are those with the least well connected longitudes, i.e. Eastern or $>$W80.}

\subsection{Contingency table and metrics}
SPARX$-$H was also tested through a contingency table using all 384 $>$M7-class flares that occurred in our event time period (up to 2017) with source longitudes between E90 and W90. The SPARX$-$H forecasts from Ch12 are compared to the high energy Ch12 SEP events from \citeA{waterfall2022hepad} to determine which flares are associated with SEPs. This results in the contingency table shown in Table~\ref{table.contingency}, where there are 39 observed $>$300\,MeV SEP events with associated flares $>$M7. We have excluded the 3 SEP events with associated flares $<$M7 from this study, and the 7 observed SEP events associated with behind limb flares from Table 3 of \citeA{waterfall2022hepad}. Due to the large uncertainties associated with source locations and flare magnitudes of behind limb events, SPARX-H does not forecast these events. From the contingency table, the probability of detection (POD), bias, false alarm ratio (FAR), accuracy and probability of false detection (POFD) are calculated, as defined in \citeA{wilks2011statistical}, and listed in Table~\ref{table.stats}. The perfect POFD and FAR scores have a value of 0, whereas the others have a perfect value of 1. From the bias score it is clear that SPARX$-$H over predicts ($>$1) SEP events for $>$M7 class flares. %In general, SPARX$-$H over forecasts high energy events however it proves to be acceptable for forecasting observed events. % False alarms and accurate event predictions are common issues in SEP forecasting and are not unique issues to our model.

\begin{table}
\caption{Contingency table for $>$M7 flares with source longitudes between E90 and W90.}
\centering
\begin{tabular}{ccc}
 & \multicolumn{2}{c}{\textbf{Observed}} \\
\textbf{Forecast} & Yes & No \\ \hline
\multicolumn{1}{c|}{Yes} & 28 & 95 \\
\multicolumn{1}{c|}{No} & 11 & 250
\end{tabular}
\label{table.contingency}
\end{table}

\begin{table}
\caption{Values determined from contingency table}
\centering
\begin{tabular}{cc}
\textbf{Metric} & \textbf{Value} \\ \hline
POD & 0.72 \\
Bias & 3.15 \\
FAR & 0.77 \\
Accuracy & 0.72 \\
POFD & 0.28
\end{tabular}
\label{table.stats}
\end{table}

%Figure~\ref{fig:sparxlocation} shows a comparison of the Ch12, Ch13 and Ch14 peak flux calculated from 18 SEP events observed by GOES-6 HEPAD compared with the peak fluxes of the historic event predicted by SPARX$-$H. All 18 historic events used occurred between 1989$-$1996 where the flare size and location for each event are input into SPARX$-$H. The GOES-6 P8 channel spans 355$-$435\,MeV, whereas SPARX$-$H predicts fluxes in HEPAD channels consistent with GOES 8-13, e.g. 350$-$420\,MeV for P8.

%Despite this slight difference in energy range for the channels, Figure~\ref{fig:sparxlocation} shows that SPARX$-$H predicts peak fluxes that are comparable to observed fluxes. The events with the best agreement between observed and predicted peak fluxes are those with source longitudes between W0$-$75, i.e. events with more well-connected source regions. There are still some Western events however where the SPARX$-$H predicted peak flux is lower than what is observed. SPARX$-$H under-predicts the peak fluxes for any SEP event observed by GOES06 with source longitudes in the East, or close to the West limb.

\section{Discussion and conclusions}
\label{sec:conclusions}
Despite the rarity of $>$300\,MeV SEP events, they are a considerable hazard to air and space travel as well as technology. The forecasting of these higher energy SEP events and reducing their potential consequences is recently becoming more necessary as space exploration and global reliance on technology increase. This is the motivation for this work, performed under the SWIMMR programme which aims to improve space weather monitoring and prediction in the UK. We have therefore extended the SEP forecasting model SPARX to proton energies above 300\,MeV. The high energy version of SPARX, called SPARX$-$H, has the capability to forecast proton fluxes in the near-Earth environment following a solar flare. The proton fluxes are calculated in 3 differential energy channels listed in Table~\ref{table.channels}, corresponding to new calibrations of GOES$-$HEPAD observations between 1984$-$2017. The availability of information on the event fluences as well as the peak fluxes provided by SPARX$-$H is valuable in understanding how long precautions need to be in place during these events. Conversely, the ability to forecast `all-clear' periods is also important.

The two major changes to the new SPARX$-$H model are the regeneration of the test particle database that feeds into SPARX (modified for higher energies), and the implementation of a new scaling for the peak fluxes using analysis of observations of higher energy SEP events. The full analysis of this dataset is given in \citeA{waterfall2022hepad}. We find improved correlations between the peak flux and flare magnitude when the events are separated according to the polarity of the IMF (i.e. A+ or A$-$). The highest correlations are seen for A+ events (0.8), however the number of high energy SEP events associated with an A+ polarity is significantly less than A$-$ events. For example, in Ch12 there are 26 A$-$ events and only 16 A+ events, dropping further for higher energy channels. Additionally, the peak flux for A$-$ events for a given flare magnitude is generally larger across all energies. The correlations are nearly all comparable to those obtained in lower energy work by \citeA{dierckxsens2015relationship}. However, the relationships between peak flux and flare intensity for A$-$ events is weaker for the highest energies. We note the smaller sample size in this work due to the less frequent occurrence of high energy events. More time (and thus more events) will enable us to further improve and refine these high energy forecasting models. However, to do so requires the availability of high energy SEP data. There are currently few instruments capable of detecting $>$300\,MeV protons and GOES$-$HEPAD provides the only long-term dataset with minimal datagaps. The current high energy particle instrument onboard GOES-R is the Space Environment In-Situ Suite (SEISS)/Solar and Galactic Proton Sensor (SGPS) covering ten differential energy channels up to 500\,MeV and one integral $>$500\,MeV channel.

SPARX$-$H uses the same trigger and input parameters as SPARX: the flare location (longitude and latitude) and magnitude. New to SPARX$-$H is the requirement of the polarity of the interplanetary magnetic field as an input. Despite only having four inputs we have seen that SPARX$-$H can reasonably predict the peak flux from historic well-connected events. SPARX$-$H predicted fluxes tend to be smaller than observed for Eastern or near-limb events. This may be due to the fact that some processes that may help SEPs from a distant active region source reach Earth, such as transport along the HCS \cite{waterfall2022modeling}, or perpendicular transport \cite{laitinen2016solar}, are not currently included in SPARX-H. It is hoped that the latter processes will be included in future forecasting models.

%SPARX$-$H uses the observed flare magnitude and peak fluxes of SEP events to scale the model count rates to physical flux values.

SPARX$-$H provides information on the time evolution of the flux, event duration, onset and fluence and is not limited to only predicting the peak flux. SPARX$-$H was tested on a set of historic events, where the predicted and observed event durations, rise times and peak fluxes were compared. Additionally, a contingency table (Table~\ref{table.contingency}) was generated based on all $>$M7 flares and $>$300\,MeV SEP events that occurred during the observed period. The probability of detection (POD) is reasonable (0.72), with SPARX$-$H able to predict the majority of observed SEP events associated with $>$M7 flares between E90 and W90, and all SEP events with associated flare locations between W25$-$W75 (POD, 1.0). However, there is also a high false alarm (FAR, 0.77) with the majority of well-connected source regions with $>$M7 flares returning a positive forecast. Generally, the majority of observed SEP events are predicted by SPARX$-$H, and the observed event parameters are most closely reproduced for events with source longitudes between W25$-$W75. Consistent verification and validation methods for SEP forecast models is a general concern, however, we have provided testing on the outputs we have available.

%Examining the predicted and observed event durations, rise times and peak fluxes reveals SPARX$-$H is best at predicting event properties for well-connected events with large (X class) flares.

SPARX$-$H currently uses one pre-generated database to produce its flux profiles. This enables a fast computation time, crucial in forecasting models. However, in future iterations of this model we hope to include other input parameters to further refine the model. For example, information on the solar wind speed or the heliospheric current sheet configuration at the time of the event may be used in future improved versions. The location of the heliospheric current sheet during high energy events has been explored by \citeA{waterfall2022modeling} and is suggested to be relevant in many large SEP and GLE events. As SEP events are also commonly associated with large CMEs, this model would ideally include information on CME speeds or width in the flux scalings as well. However, as is seen in \cite{waterfall2022hepad}, the correlations between CME speeds and SEP peak fluxes at high energies are very poor. Additionally, unlike for solar flares, there is currently no real-time information on CME properties available. Future iterations of this model will benefit from the inclusion of information on accompanying CMEs.

\section*{Data availability statement}
This manuscript has used the SEPEM reference dataset version 3 \cite{dataset}, which has been extended to high energies using re-calibrated HEPAD data \cite{raukunen2020very}. Information on the polarity of the interplanetary magnetic field was obtained from the Wilcox Solar Observatory (WSO), currently operated by Stanford University
with funding provided by the National Science Foundation. WSO data is accessible from \citeA{murdin2000wilcox}. Figures were created using Matplotlib version 3.3.2 \cite{hunter2007matplotlib}.

%This manuscript has used the SEPEM reference dataset version 3 \cite{dataset}, which has been extended to high energies using re-calibrated HEPAD data \cite{raukunen2020very}. The data was originally obtained from GOES observations which was cleaned, background subtracted and cross-calibrated \cite{sandberg2014, rodriguez2017}. The original data is available from: \url{satdat.ngdc.noaa.gov/sem/goes/data/avg/}. Neutron monitor data was obtained from the GLE database \url{https://gle.oulu.fi}, managed and hosted by the Oulu Cosmic Ray Station at the University of Oulu, Finland \cite{usoskin2020revised}. Information on the heliospheric current sheet was obtained from the Wilcox Solar Observatory (WSO), currently operated by Stanford University with funding provided by the National Science Foundation. WSO data is accessible via \url{http://wso.stanford.edu}. Information on existing SEP events was obtained from the NOAA Space Environment Services centre list at: \url{https://umbra.nascom.nasa.gov.SEP/}. Figures were created using Matplotlib version 3.3.2 \cite{hunter2007matplotlib}, available from \url{https://matplotlib.org}.

\begin{acknowledgments}
C.O.G.~Waterfall and S.~Dalla acknowledge support from NERC via the SWARM project, part of the SWIMMR programme (grant NE/V002864/1).
SD acknowledges support from the UK STFC through grants ST/R000425/1 and ST/V000934/1 and from the International Space Science Institute through funding of the International Team on `Solar Extreme Events: Setting up a paradigm'. We acknowledge the use of data from Wilcox Solar Observatory data in this study.
\end{acknowledgments}

%%    This version assumes use of bibtex with the jswsc.bib file being present
%%    If your bib file has a different name you need to change the following line

\bibliography{main}

%\end{linenumbers}

\end{document}